 \documentclass[final,3p,times]{elsarticle}

\usepackage[colorlinks=true, linkcolor=blue, urlcolor=blue]{hyperref}
\usepackage{doi}

\usepackage{amssymb}
\usepackage{amsmath}

\biboptions{numbers,sort&compress}

\journal{Physica A: Statistical Mechanics and its Applications}

\begin{document}

\begin{frontmatter}



\title{Intermediate Interaction Strategies for Collective Behavior}


\author[1,2]{Yuto Kikuchi\corref{cor1}} 
\ead{00kyuto@gmail.com} 
\author[3,4]{Mayuko Iwamoto} 

\affiliation[1]{organization={Graduate School of Integrated Sciences for Life, Hiroshima University},
            city={Higashihiroshima}, 
            state={Hiroshima},
            country={Japan}}

\affiliation[2]{organization={Nagoya University, Graduate School of Medicine},
            city={Nagoya},
            state={Aichi},
            country={Japan}}

\affiliation[3]{organization={Future University Hakodate},
            city={Hakodate},
            state={Hokkaido},
            country={Japan}}

\affiliation[4]{organization={Meiji Institute for Advanced Study of Mathematical Sciences(MIMS), Meiji University},
            city={Nakano},
            state={Tokyo},
            country={Japan}}

\cortext[cor1]{Corresponding author} 

\begin{abstract}
From bird flocks and fish schools to migrating cell sheets, collective motion is a ubiquitous biological phenomenon that inspires quantitative modeling through self-propelled particle (SPP) frameworks.
Conventional SPP models prescribe either distance-based (metric) or rank-based (topological) interactions; however, empirical studies indicate that real groups may blend both types of interaction. 
Motivated by this graded perception, we introduce a new three-dimensional SPP model in which metric and topological alignments act simultaneously and are weighted by a single tunable mixing parameter called the interaction parameter.
Large-scale simulations spanning a wide ranges of interaction parameters and densities revealed rich dynamics. 
Even when the global order parameter is low, cluster-level analysis with HDBSCAN shows that particles self-organize into several spatially distinct but internally well-aligned sub-flocks, exposing a hidden layer of order.
Most importantly, an intermediate balance in which metric and topological cues contribute almost equally maximizes the global order parameter and markedly improves robustness to density variations.
Numerical experiments supported by linear stability analysis demonstrate that activating metric and topological interactions in concert bridges the traditional modeling dichotomy and furnishes a more adaptive and resilient framework for collective motion.
Therefore, the proposed model therefore provides a versatile platform for exploring mixed-interaction effects in biological and engineered multi-agent systems.
\end{abstract}




\end{frontmatter}



\section{Introduction}
\label{sec1}
Collective behavior, whether exhibited by living or non-living entities, is ubiquitous in nature \cite{bib11}\cite{bib21}.
Ecological examples include the schooling of fish\cite{bib21}\cite{bib1}\cite{bib22}, flocking of birds \cite{bib23}\cite{bib24}\cite{bib10}\cite{bib9}, and collective cell migration\cite{bib25}\cite{bib26}\cite{bib27}. 
Numerous studies have explored these biological collective phenomena, resulting in a wide variety of mathematical formulations.
Among these, self-propelled particle (SPP) models have played a key role. 
Numerous extensions exist, such as a visual-field-based model of fish schooling \cite{bib1} and a framework for understanding collective decision-making during the landing phase of bird flocks\cite{bib23}.  
Comparable progress has been made in robotic control laws, such as robot-herding algorithms inspired by sheep-dog systems\cite{bib14} and a one-dimensional car-following model that reproduces single-lane traffic dynamics\cite{bib2}.
There are also studies that drawing inspiration from pedestrian flow and capture the essential control mechanisms for achieving multi-objective tasks\cite{bib15}\cite{bib16}\cite{bib17}\cite{bib18}.
The archetypal SPP framework is the Vicsek model \cite{bib3} \cite{bib4}.
The Vicsek model generalizes Reynolds Boids system---which employs metric-neighbor interactions to describe flocking \cite{bib5}---to a broader class of situations. 
In the Vicsek model, each particle moves at a constant speed while experiencing a local alignment interaction that tends to match its direction of motion with nearby particles, along with stochastic angular noise. 
By varying the noise strength, the particle ensemble shifts from a disordered state to a collectively aligned state, which constitutes a phase transition interpreted as spontaneous symmetry breaking in a non-equilibrium system\cite{bib2}.

Empirical studies have shown that real biological collective behavior can be reorganized into different group morphologies governed by alternative interaction rules\cite{bib9}\cite{bib29}. 
Investigations based on the Vicsek model, which examines interaction mechanisms within collective behavior, have greatly facilitated the revelation of these processes.
Recent studies contrast the two principal definitions of interaction: metric interactions\cite{bib3}\cite{bib4}, determined by the physical distances between individuals, and topological interactions\cite{bib9}\cite{bib10}\cite{bib28}, determined by nearest-neighbor rank. 
Metric interactions promote density-dependent grouping, whereas researchers have explored a different type of interaction known as a topological interaction\cite{bib6}.
Examples include selecting neighbors according to the topological distance derived from the Voronoi tessellation\cite{bib12}\cite{bib13} or interacting with a fixed number of nearest neighbors ranked by geometric distance\cite{bib10}.
An analysis that extended three-dimensional positional data of jackdaw (\textit{Corvus monedula}) flocks using a three-dimensional Vicsek model revealed context-dependent interaction rules\cite{bib9}.
Specifically, jackdaws relied on metric interactions during transit flocks but switched to topological interactions during mobbing flocks, thereby adapting group morphology to behavioral context.
Hybrid models that switch between metric and topological rules have been proposed and proven effective in reproducing the scale-free correlation patterns characteristic of biological flocks\cite{bib7}\cite{bib8}.
They modeled the internal noise by allowing the interaction rule itself to switch randomly from metric to topological modes.

In contrast to the discrete-switching interactions assumed in conventional models, observations indicate that real animals, particularly birds and fish, perceive distance, visual coverage, and group size in graded and continuous quantities\cite{bib10}.
Thus, the range of interaction between individuals most likely varies along a perceptual-cognitive continuum rather than switching between two discrete modes. 
Accordingly, it is natural to assume that agents weigh their interactions, responding strongly to nearby neighbors while being influenced only slightly by those farther away.
Embedding such perceptual ambiguity into a modelling framework should yield governing rules that are both more realistic and robust. 
Accordingly, we formulate a model that spans the intermediate phenomena observed between metric and topological interactions, and examine how their concurrent influence shapes the collective order and its morphological transitions.
This intermediate perspective incorporates biological flexibility and fuzzy perceptual boundaries that are absent from previous switching approaches, thus providing a stronger theoretical basis for inferring control laws in living groups.

In this study, we introduce a novel agent-based framework in which two distinct interaction mechanisms operate simultaneously.
We developed an extended three-dimensional Vicsek model that smoothly changes between metric and topological interactions using a single control parameter, thereby making their intermediate behaviors explicit. 
Through numerical simulations, we investigated how the modulation of local interactions regulates the emergence and morphology of collective order. 
The proposed framework constitutes a natural theoretical extension of conventional agent-based models and deepens our understanding of collective dynamics.
It also enables robust descriptions of biological swarms under perceptually ambiguous conditions where switching approaches are inadequate.
By incorporating the biological flexibility and perceptual ambiguity characteristic of real organisms that are absent in conventional models, our framework enhances the generality and adaptability of control-oriented mathematical descriptions.

\begin{figure}[t]
\centering
\includegraphics{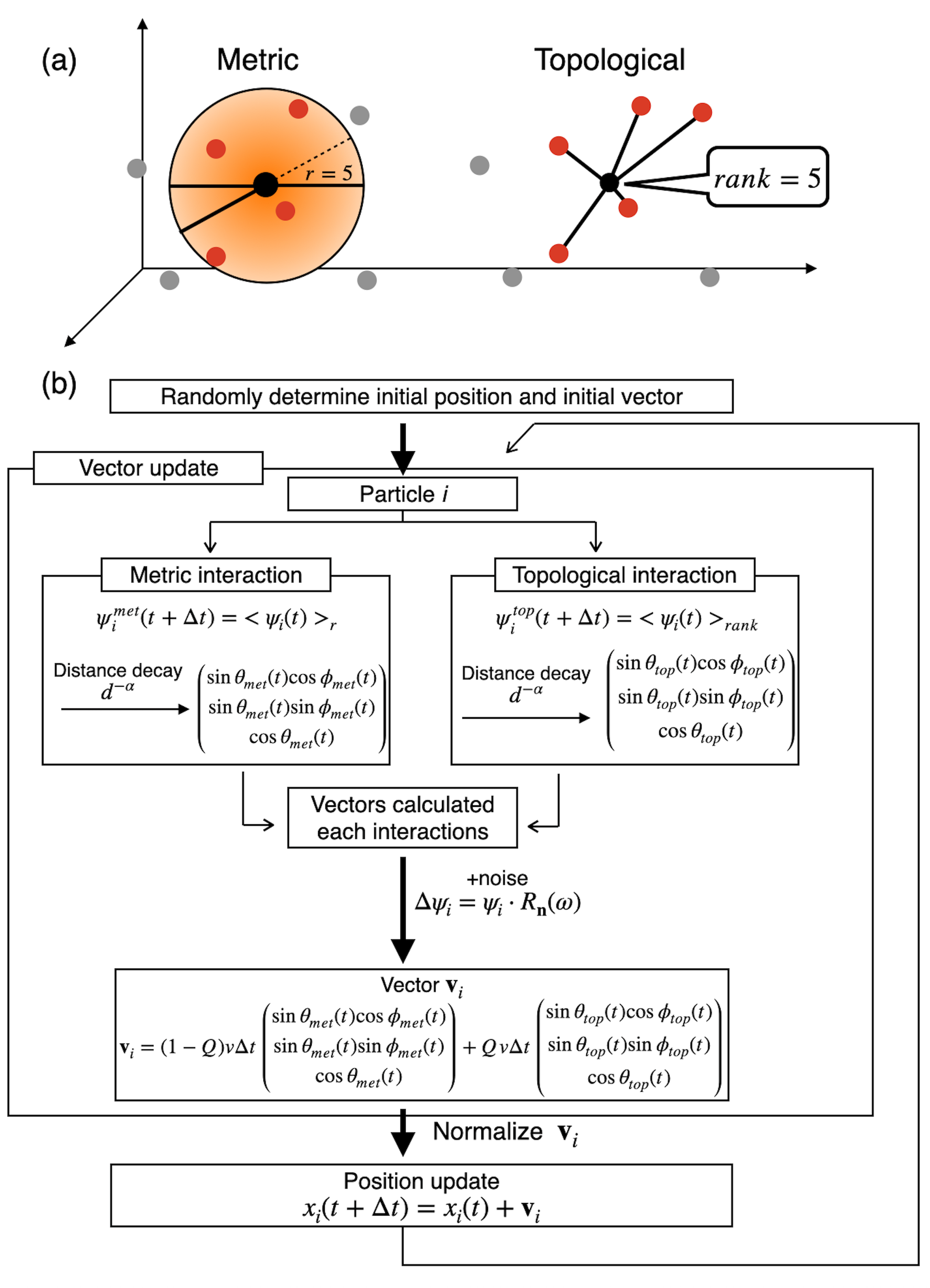}
\caption{Illustration of each interaction type and a flowchart of the algorithm for the 3D-KI model a) Schematic illustration of the two interaction rules.
The focal particle is shown in black and its interacting neighbors in red.
(Left) Metric interaction : 
The focal particle averages the direction of motion of all neighbors located within a spherical region of radius $r$ (orange shading ; $r=5$ in this example).
(Right) Topological interaction : 
The focal particle averages the directions of motion of its rank-nearest neighbors (here, rank$=5$).
In both cases, neighbor contributions are distance-weighted so that influence decays with separation.
(b) Algorithm of the proposed 3D-KI model.
Particles are initialized with random positions and directions of motion. At each time step, every particle calculates the metric and topological alignment vectors, combines them according to the interaction parameter $Q$,
and updates its direction of motion accordingly.}\label{fig1}
\end{figure}

\section{Methods}
\label{sec2}

\begin{figure}[t]
\centering
\includegraphics{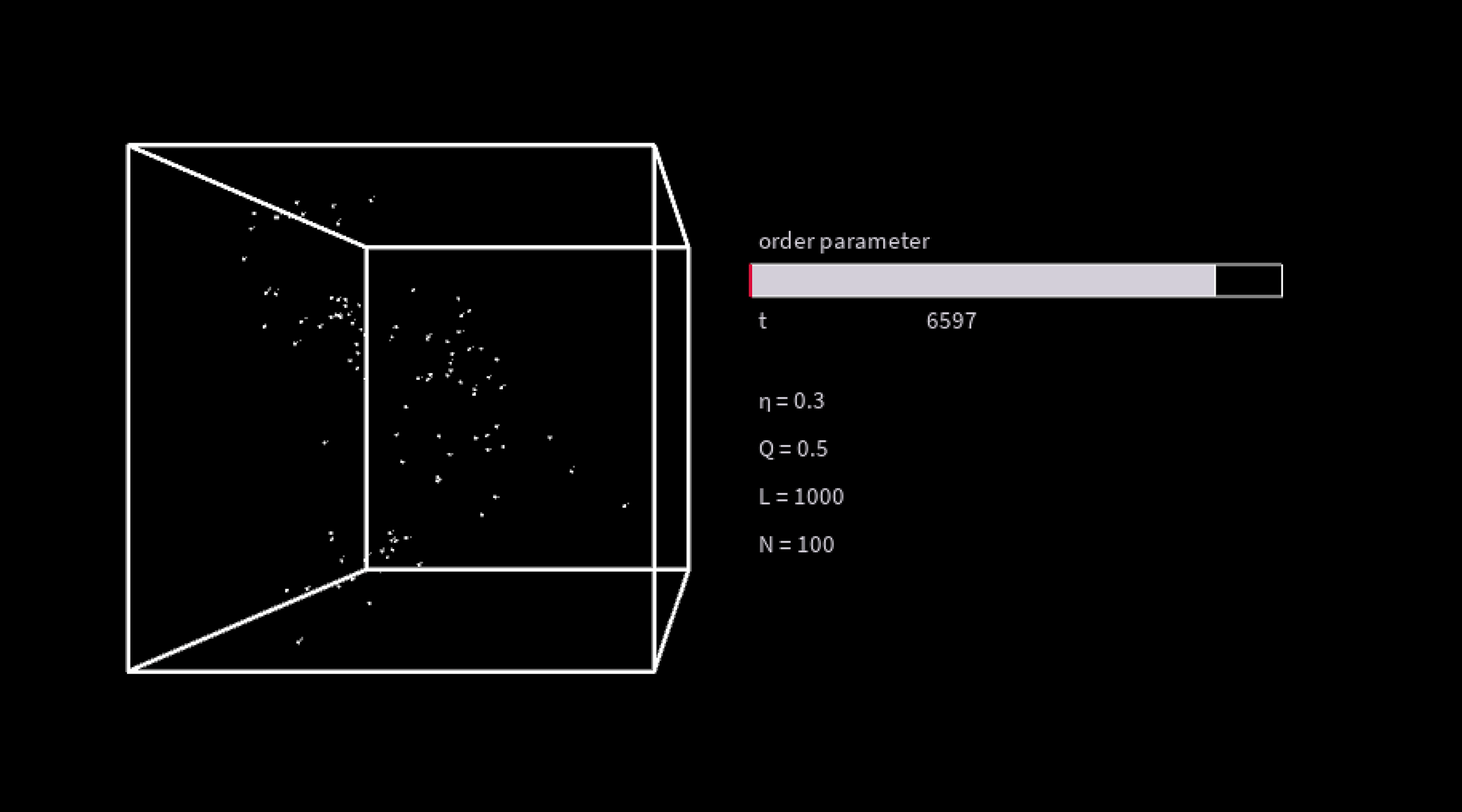}
\caption{Time snapshot of a 3D self-propelled particle (SPP) simulation with periodic boundary conditions. 
The visualization includes a 3D box frame to indicate system boundaries and a progress bar to show the value of the order parameter $v_a$.}\label{figSimu}
\end{figure}

In this study, we introduce a 3D Kinematic Metric-Topological Intermediate-interaction Model (3D-KI model), which is a novel agent-based framework in which two distinct interaction mechanisms operate simultaneously.
Consider a particle whose position is represented by ${\bf x}=(x,y,z)$ in Cartesian coordinates and $(\gamma,\theta,\phi)$ in spherical coordinates.
The anguler parameters are defined as follows:
\begin{align}
\theta&=\arccos\frac{z}{\gamma} \\
\phi&=\frac{1}{1+\exp(-ay)}\arccos\frac{x}{\sqrt{x^2+y^2}}\ ,
\end{align}
where $\gamma=\sqrt{x^2+y^2+z^2}$ denotes the radial distance from the origin.
Here, $\phi$ denotes the azimuthal angle in the $x-y$ plane, and $\theta$ is the polar angle measured from the $z$-axis.
To isolate the effects of the interaction scheme, we assumed that all particles moved at a constant speed.
Consequently, the position vector of particle $i$ after a small time increment $\Delta t$ is determined solely by this fixed magnitude and the direction of particle $i$ at time $t$.

A metric interaction is defined by a geometric neighborhood: each particle aligns its direction of motion with the average direction of all other particles located inside a spherical region of radius $r$ centered on itself (Left panel in Fig.\ref{fig1}a). 
Because the interactions depend on this fixed distance, they are strongly influenced by the total number of particles, system size, and local density.
Once a neighbor enters the interaction area, the alignment persists until random fluctuations or noise carry that neighbor outside the radius.
In our model, a topological interaction is realized by aligning each particle with the average direction of motion of its ranked nearest neighbors, where the rank is a prescribed topological parameter (Right panel in Fig.\ref{fig1}a).
A key difference emerges: the metric interaction involves “all individuals within radius $r$,” so the number of interacting neighbors grows with local density, whereas the topological interaction maintains a constant number of interacting neighbors and is effectively insensitive to density variations.

A small kinematic ambiguity was introduced as an external perturbation using the rotation-matrix form of Rodrigues’ formula.
Given an arbitrary unit vector ${\bf n}=\left( n_1,n_2,n_3 \right)$ that defines an axis through the origin, the matrix:
\begin{equation}
 R_n(\omega)=
 \begin{bmatrix}
   n_1^2(1-\cos(\omega))+\cos(\omega) & n_1n_2(1-\cos(\omega))-n_3\sin(\omega) & n_1n_3(1-\cos(\omega))+n_2\sin(\omega) \\
   n_1n_2(1-\cos(\omega))-n_3\sin(\omega) & n_2^2(1-\cos(\omega))+\cos(\omega) & n_2n_3(1-\cos(\omega))-n_1\sin(\omega) \\
   n_1n_3(1-\cos(\omega))+n_2\sin(\omega) & n_2n_3(1-\cos(\omega))+n_1\sin(\omega) & n_3^2(1-\cos(\omega))+\cos(\omega) 
 \end{bmatrix}
\end{equation}
$R_n(\omega)$ rotates any vector by an angle $\omega$ around the axis, thereby providing the desired noise term for each time step.
Multiplying $R_n(\omega)$ by the unit velocity vector of particle $i$, $\psi_i(t)$, yields a noise-perturbed direction of motion.
In this study, the rotation angle $\omega$ was defined as follows: 
\begin{equation}
\omega = \eta\cdot\omega'
\end{equation} 
where $\eta\in \mathbb{R}^+ ,\ \omega'\in[-\pi,\pi]$.
The introduction of a noise parameter $\eta$ makes it possible to control the magnitude of random fluctuations, thereby embedding spatial noise into the model.

In our model, we introduce an interaction parameter $Q$ that quantifies the relative contributions of metric and topological interactions.
When $Q=0$, the update rule reduces to a purely metric interaction model, whereas when $Q=1$, it yields a purely topological interaction model.
For an intermediate value $0<Q<1$, the mean vectors obtained from the metric and topological neighbors are linearly combined in the ratio specified by $Q$.
Therefore, we captured the dynamics that emerge when both interactions occur simultaneously.
In three-dimensional space, the position of particle $i$ after a small time step $\Delta t$, ${\bf x}_i(t+\Delta t)$, and its unit vector $\psi_i(t)$ at the same instant are given by
\begin{equation}
{\bf x}_i(t+\Delta t)={\bf x}_i(t)+(1-Q)v\Delta t
\begin{pmatrix}
\sin\theta_{met}(t)\cos\phi_{met}(t) \\
\sin\theta_{met}(t)\sin\phi_{met}(t) \\
\cos\theta_{met}(t)
\end{pmatrix}+Qv\Delta t
\begin{pmatrix}
\sin\theta_{top}(t)\cos\phi_{top}(t) \\
\sin\theta_{top}(t)\sin\phi_{top}(t) \\
\cos\theta_{top}(t)
\end{pmatrix}
,
\end{equation}
\begin{equation}
\psi_i(t+\Delta t)=\left<\psi_i(t)\right>+\Delta\psi_i\ ,
\end{equation}
where
\begin{equation}
\left<\psi_i(t)\right>=\arctan\left(\frac{\left<\sin\psi_i(t)/d^{\alpha}\right>}{\left<\cos\psi_i(t)/d^{\alpha}\right>}\right)\ ,
\end{equation}
and ${\bf x}_i$ is the position of particle $i$, $\theta_{met}$ and $\phi_{met}$ are angular parameters calculated by metric interaction; and $\theta_{top}$ and $\phi_{top}$ are angular parameters calculated by topological interaction.
Here, $\left<A\right>$ is the average value of A.

Qur model introduces a distance-dependent weighting scheme.  
When computing the mean direction of motion, the contribution of neighbor $j$ is multiplied by $d^\alpha_{j}$, where $d_{j}$ is the interparticle distance, and $\alpha\in {\mathbb R}^+$ is a tunable exponent.
This inverse-power weighting allows nearer neighbors to influence the focal particle more strongly than those farther away.
$\left<\frac{\sin\psi_i}{d^{\alpha}}\right>$ and $\left<\frac{\cos\psi_i}{d^{\alpha}}\right>$ are given by:
\begin{equation}
\begin{cases}
\left<\frac{\sin\psi_i}{d^{\alpha}}\right>=\frac{1}{M}\Sigma_{j=0}^{M}\frac{\sin\psi_j(t)}{d_j^\alpha}\\
\left<\frac{\cos\psi_i}{d^{\alpha}}\right>=\frac{1}{M}\Sigma_{j=0}^{M}\frac{\cos\psi_j(t)}{d_j^\alpha}
\end{cases}
\end{equation}
Here, $M$ is the number of neighbors contributing to the interaction, and $\alpha$ controls how strongly the interparticle distance is considered. 
When $\alpha=0$, the model entirely excludes distance-dependent attenuation.

To monitor the temporal evolution of the global collective state, we introduced the order parameter $v_a$:
\begin{equation}
v_a(t)=\frac{1}{N\cdot v}\left|\sum^{N}_{i=1}{\bf v}_i\right|
\end{equation}
the velocity of each particle is given by ${\bf v}_i=v\psi_i(t)$, where $v$ is the constant speed and $\psi_i(t)$ is the unit vector.
The order parameter $v_a(t)$ lies in the interval $\left[0,1\right]$ and quantifies the degree of directional alignment at step $t$.
It reaches $v_a=1$ when all $N$ particles share an identicial direction of motion and approaches $v_a=0$ when the ensemble becomes increasingly disordered. 

In this study, periodic boundary conditions were employed to simulate an environment where a particle group can move without external spatial constraints.
This assumption reflects the nature of many biological systems where collective behavior typically emerges in open or fluid environments rather than within closed or confined spaces.
For example, flocks of birds, schools of fish, and migrating cells often form and maintain group-level coordination without being restricted by rigid external boundaries.
Periodic boundaries allow us to avoid artificial edge effects that might otherwise distort a group’s spatial organization and alignment.
Thus, the dynamics observed in our simulations more accurately represent the intrinsic properties of the system, independent of finite size or boundary-induced artifacts.
Figure \ref{figSimu} is a snapshot from a simulation of the proposed model. 
Point-like (zero-volume) particles were rendered as white dots, with a wireframe cube indicating the simulation domain. 
The right-hand panel reports the instantaneous order parameter $v_a$, simulation time $t$, noise parameter $\eta$, interaction parameter $Q$, domain side length $L$, and particle count $N$. 
The visualisation was created in Processing 4.3.4 (see Supplementary Movie S1).

\section{Results}
\label{sec3}
\begin{figure}[t]
\centering
\includegraphics{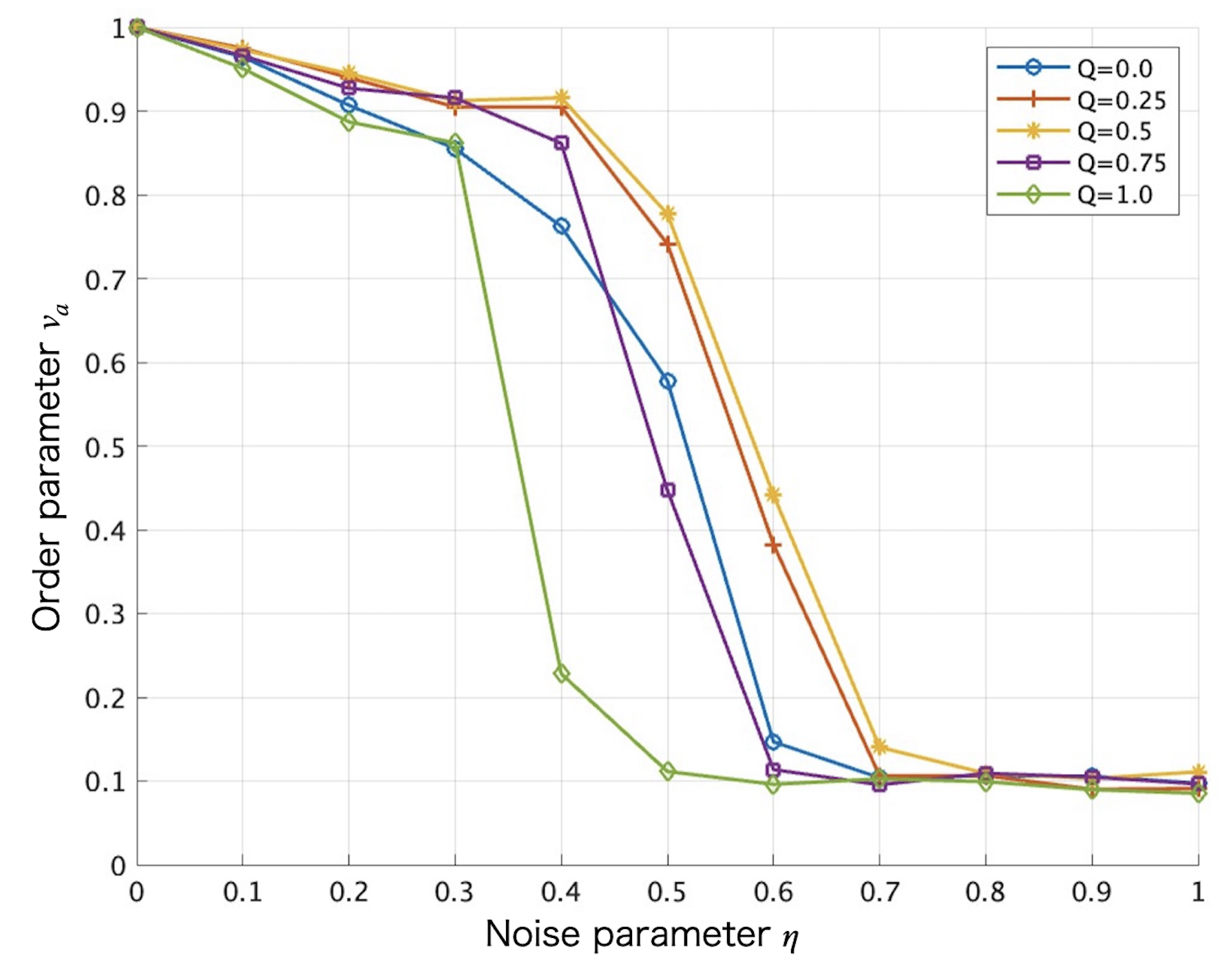}
\caption{Dependence of the order parameter $v_a$ on the noise parameter $\eta$. 
Simulations were conducted with the interaction parameter $Q$ set to 0.0(Metric), 0.25, 0.5, 0.75, and 1.0(Topological).
The number of particles was 100, and the domain size was set to 400×400×400 ($\rho=1.6\cdot 10^{-6}$).
Each data point represents the average of 50 independent simulations with randomized initial conditions.
In our model, a sharp decrease in the order parameter under topological interaction---previously reported in a 2D model by reference\cite{bib6}---was also observed.
}\label{fig2}
\end{figure}
In this section, we descrive the properties of our 3D-KI model.
First, we compare the values of the order parameter $v_a$ obtained for different values of the interaction parameter $Q\in [0,1]$ with those from the purely metric model ($Q=0.0$) and the purely topological model ($Q=1.0$).
The invariant parameters were set as follows: the number of particles $N$ was fixed at 100, the three-dimensional domain was a cube with a side length of $L=400$, the radius $r$ of the spherical region for the metric model was 100, the number of topological neighbors (rank) was set to 6, and the distance decay parameter $\alpha$ was fixed at 1.0.
We defined the density as $\rho=N/L^3$(in this case($N=400$) and $\rho=16\cdot 10^{-7}$). 
Fig. \ref{fig2} shows the relationship between the order parameter $v_a$ and the noise parameter for each value of $Q$. 
For each condition, we performed 50 simulations of 10,000 steps with randomized positions for each individuals and calculated the mean of the final values of $v_a$ (for details, see Supplementary Fig. S1).
For all interaction parameters, including the metric, topological, and intermediate values of $Q$, the order parameter $v_a$ decreased as the noise parameter $\eta$ increased.
These results indicate that even in ambiguous environments where both types of interactions coexist, an increase in noise consistently leads to a decrease in polarization. 
This suggests the presence of a noise-induced phase transition in the system.
Moreover, we found that the average order parameter increases from the metric side and decreases toward the topological side, reaching a peak at $Q=0.5$.
In the range from $Q=0.0$ to $Q=0.5$, the results indicate that introducing elements of topological interaction enhances the average value of the order parameter, particularly under low-noise conditions (noise parameter $\leq 0.6$).
In the range from $Q=0.5$ to $Q=1.0$, strengthening the topological interaction resulted in a sharper response of the order parameter to noise.

\begin{figure}[t]
\centering
\includegraphics{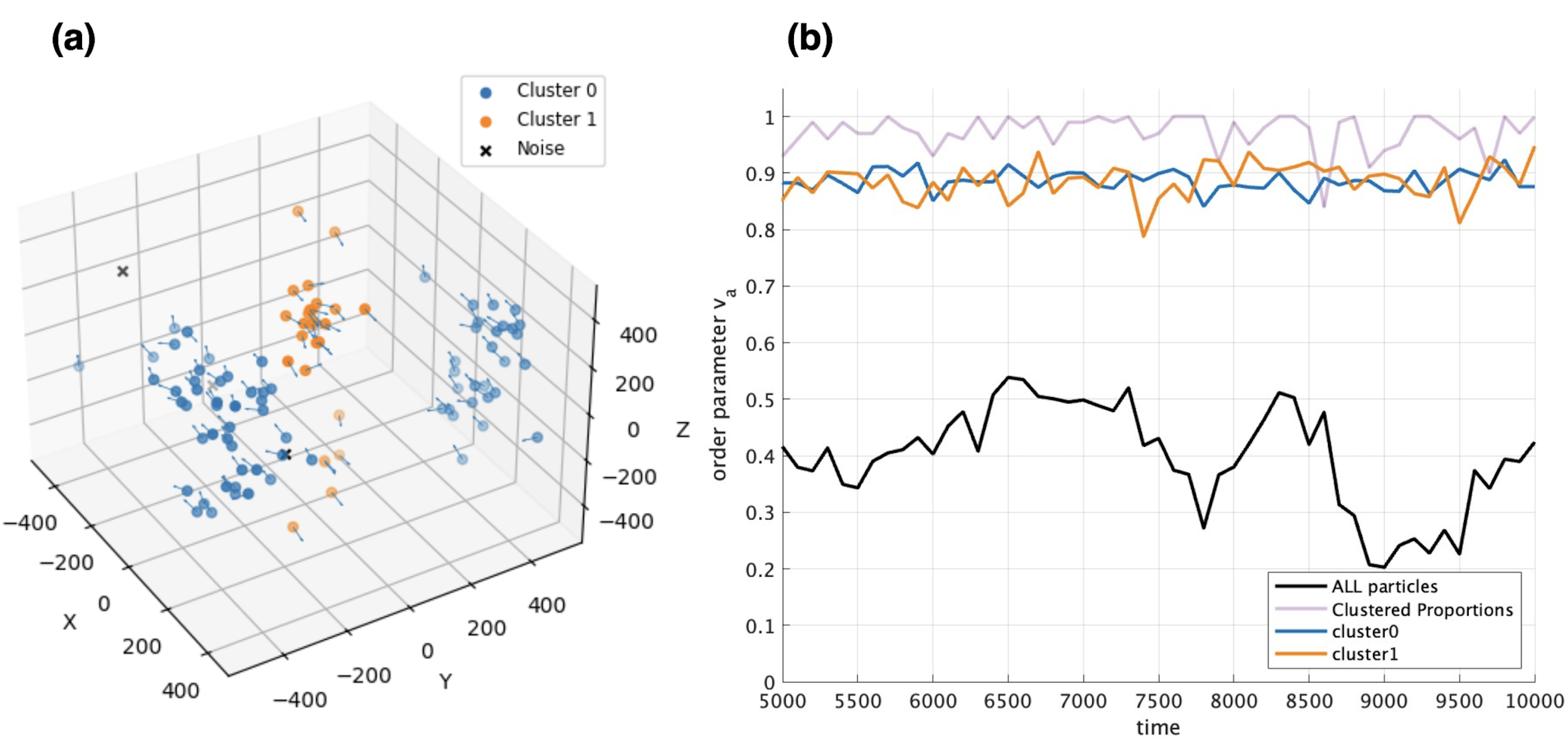}
\caption{(a)Clustering using HDBSCAN revealed that in simulations exhibiting low grobal order parameters, particles did not form a single cohesive group but instead split into multiple distinct clusters. Particles labeled as "noise" represent those not belonging to any cluster.
(b)Comparison between the order parameter computed over all particles and those calculated for individual clusters, in a state where more than 90\% of particles were successfully clustered. 
While the global order parameter remains low, indicating a lack of global alignment, the cluster-wise order parameters are around 0.9, suggesting that several highly aligned clusters coexist. 
Simulation parameters: $\rho=1.0\cdot10^{-7}$, $N=100$, $Q=1.0$, $\eta=0.3$.}\label{fig6}
\end{figure}
Previous studies indicated that self-propelled particle (SPP) models governed by topological interactions often fragment into several distinct clusters within a simulation domain\cite{bib6}\cite{bib32}. 
Motivated by these reports, we first asked whether the same multi-cluster behavior would arise in our model (see Supplementary Movie S2).
Because a low global order parameter can conceal well-aligned structures at smaller scales, we supplemented this metric with an explicit clustering analysis. 
We applied Hierarchical Density-Based Spatial Clustering of Applications with Noise (HDBSCAN)\cite{bib19}\cite{bib20} to detect clusters of arbitrary shapes and densities while automatically flagging outliers. 
Interparticle distances were evaluated as the minimum separations under periodic boundary conditions, and clusters with fewer than 10 particles were discarded. 
Analyses were restricted to frames in which over $90\%$ of particles were assigned to clusters; for each frame, we compared the order parameter of every cluster with the global value. 
Figure \ref{fig6}(a) shows that our model likewise produces the multiple local clusters, as revealed by the HDBSCAN algorithm, consistent with previous studies.
Even when the global order parameter indicated a disordered state, the intra-cluster order parameter consistently exceeded 0.8 (Fig.\ref{fig6}(b)).
This indicates that, even when the global order parameter suggests a disordered state, the particles within each cluster remain highly aligned.
This implies that the introduction of topological interactions leads to the formation of spatially separated but internally coherent sub-flocks, whose collective orientations are nearly perfectly synchronized.

\begin{figure}[t]
\centering
\includegraphics{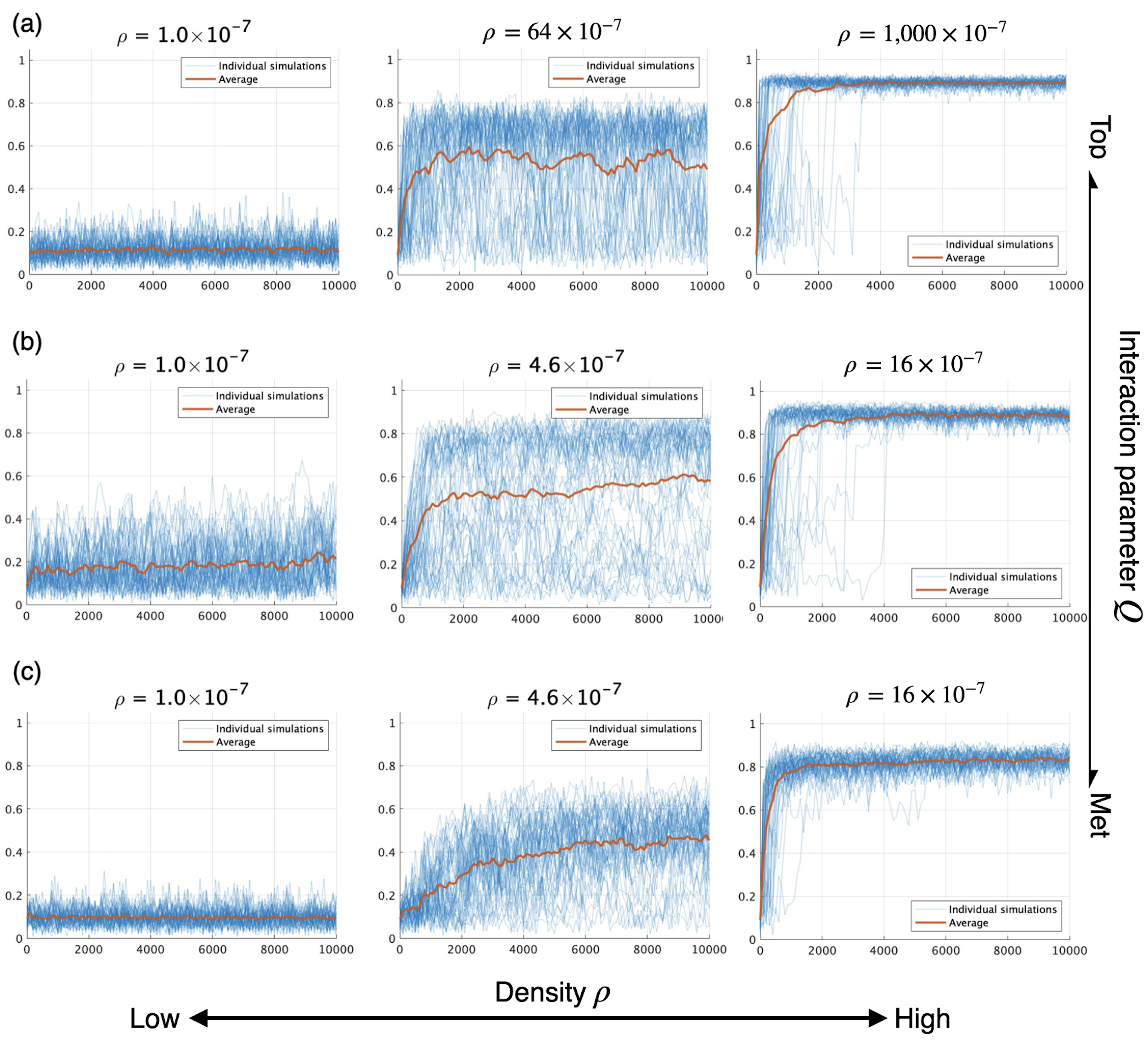}
\caption{Time evolution of the average order parameter $v_a$ and visualization of its variance across 50 simulation trials, as presented in Fig.\ref{fig3}. 
The blue lines represent individual simulations with different initial conditions, while the red line indicates their average. 
For each interaction parameter $Q$, three representative simulations with notably different variances are selected.
(a) Simulations under $Q = 1.0$ (topological only), from left to right: $\rho = 1.0\cdot10^{-7}, 64\cdot10^{-7}, 1,000\cdot10^{-7}$.
(b) Simulations under $Q = 0.7$, from left to right: $L = \rho = 1.0\cdot10^{-7}, 4.6\cdot10^{-7}, 16\cdot10^{-7}$.
(c) Simulations under $Q = 0.0$ (metric only), from left to right: $L = \rho = 1.0\cdot10^{-7}, 4.6\cdot10^{-7}, 16\cdot10^{-7}$.}\label{fig5}
\end{figure}
To obtain a picture of the characteristic dynamics produced by our model, we first extracted the time evolution of the average order parameter $v_a$ from 50 simulations and compared it across density $\rho$ and interaction parameter $Q$.
Figure \ref{fig5} presents representative time evolutions of the order parameter $v_a$ for parameter combinations selected to highlight the most characteristic behaviours observed in our simulations.
We examine three interaction regimes---(a)$Q=1.0$ (topological), (b)$Q=0.7$ (mixed), and (c)$Q=0.0$ (metric).
For each regime, three densities $\rho$ are chosen so that the variance differences among these densities are maximized.
In esch panel, the thin blue curves represent the 50 individual simulations, the bold red curve indicates the average, and the density increases from left to right.
Across all values of $Q$, the system remains weakly aligned and highly fluctuating at low density, undergoes a rapid increase in $v_a$ once a mid-density threshold is crossed, and maintains a strongly ordered state at high density.
A comparison of the center column reveals a systematic change in the steepness of this transition: the purely metric interaction ($Q=0.0$) produces the smallest increase, the mixed case ($Q=0.7$) is steeper, and the purely topological interaction ($Q=1.0$) is the steepest.
In the high-density column, the topological interaction yields larger run-to-run variability in the time required for $v_a$ to converge, yet it exhibits the smallest variance once convergence is achieved.
Conversely, the metric interaction converges at more uniform times but retains a comparatively larger post-convergence variance.
Taken together, these findings suggest a complementary division of labor between the two interaction types: metric interaction accelerates the formation of a single cohesive cluster, whereas topological interaction enhances the long-term stability of the cluster’s aligned state.

\begin{figure}[t]
\centering
\includegraphics{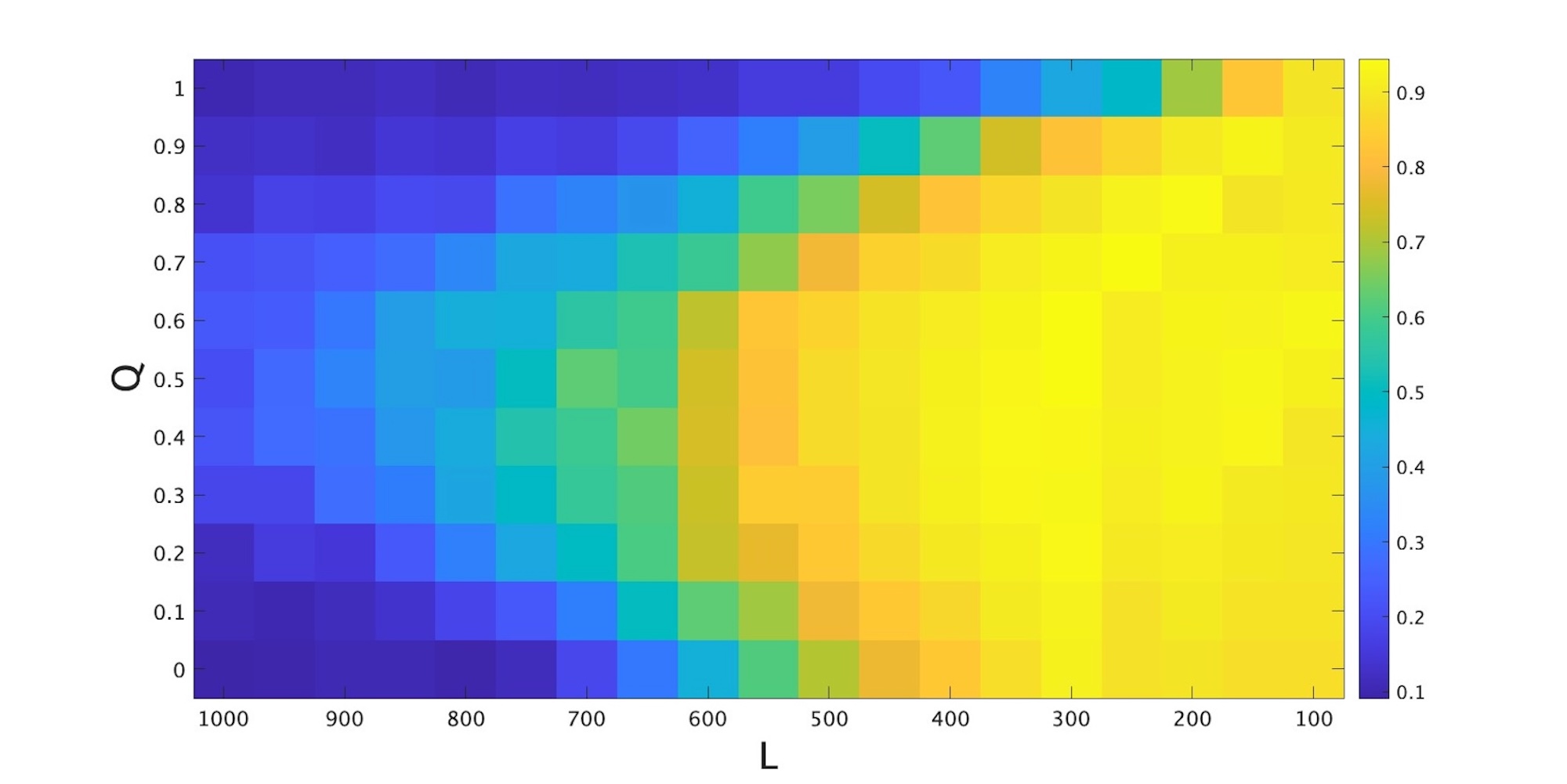}
\caption{ 
Heatmap of the converged values of the order parameter as a function of the interaction parameter $Q$ and particle density. 
Each value represents the average over 50 simulations with randomly initialized conditions, each running for 10,000 time steps. 
Density was controlled by fixing the number of particles at 100 and varying the system size from 1,000 to 100 in increments of 50. 
This adjustment of $L$, corresponds to density transitions between $1.0\cdot10^{-7}$ and $1,000\cdot10^{-7}$.
Lower density is on the left, and higher density is on the right.}\label{fig3}
\end{figure}
Next, we fixed the noise parameter $\eta$ to 0.4 and investigated how variations in particle density $\rho$ affect the order parameter $v_a$ depending on the interaction parameter. 
Specifically, we varied the interaction parameter $Q$ from 0.0 to 1.0 in increments of 0.1 and conducted simulations for 10,000 time steps under different density conditions for each value of $Q$.
The aim of this analysis was to clarify how different interaction schemes---from metric- to topological-based—respond to changes in particle density $\rho$ in terms of their collective ordering behavior.
The particle density was adjusted by changing the side length $L$ of the cubic domain from 100 to 1,000 while maintaining a constant number of particles. 
This setup effectively altered the system's volume and, hence, the particle density. 
Accordingly, the density varied in the range of $1.0\cdot10^{-7}$ to $1000\cdot10^{-7}$ as the domain size changed.
We systematically evaluated the response of the order parameter under a wide range of density conditions.
Figure \ref{fig3} reveals that the final value of the order parameter increases with density across all values of the interaction parameter $Q$.
In addition, the change in the order parameter appears to be symmetrically distributed around $Q=0.5$, suggesting the presence of a parabolic critical boundary with an apex at $Q=0.5$.

\begin{figure}[t]
\centering
\includegraphics{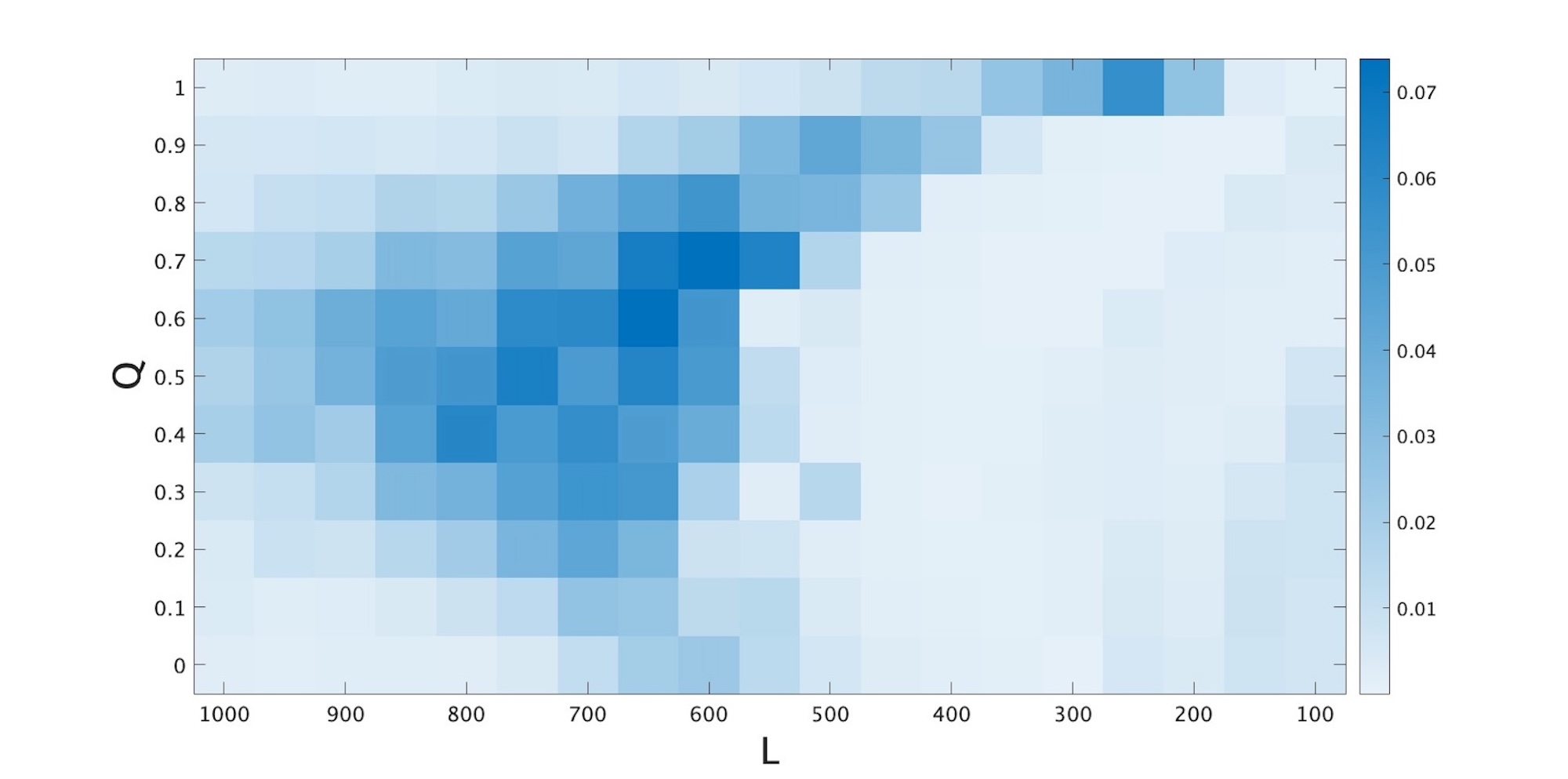}
\caption{Heatmap of the variance of the order parameter with respect to the interaction parameter $Q$ and system density. The vertical and horizontal axes are set as in Fig. \ref{fig3}, with the vertical axis representing the interaction parameter and the horizontal axis representing the edge length of the cubic domain. 
The variance was calculated using the same dataset employed in Fig. \ref{fig3}, without performing additional simulations. 
Relatively high variance values are observed in regions where the mean order parameter in Fig. \ref{fig3} ranges from 0.3 to 0.7.
In addition, greater variance is observed under topological interaction conditions compared to metric ones.
}\label{fig4}
\end{figure}
Beyond the quantitative analysis of the collective states described above, we calculated the variance of the order parameter obtained under identical conditions and visualized it as a heatmap (Fig. \ref{fig4}). 
This enabled a spatial evaluation of the stability and magnitude of fluctuations in the ordered states.
Using the mean values(Fig. \ref{fig3}) and variance(Fig. \ref{fig4}) of the order parameter, we assessed not only the degree of order in the collective behavior under each condition but also its reproducibility and robustness.
We found that the variance was particularly high near the boundary region where the order parameter shown in Fig. \ref{fig3} ranged from $Q\in [0.3,0.7]$.
Furthermore, a clear tendency was observed in which the conditions with stronger topological interactions ($Q\in [0.6,1.0]$) exhibited greater variance than those with stronger metric interactions ($Q\in [0.0,0.4]$); the average variances were 0.0173 and 0.0126 in the former and latter cases, respectively.

\section{Discussion and conclusion}
\label{sec4}
In this study, we constructed a novel self-propelled particle (SPP) model in which two interaction types act concurrently and elucidated the resulting control characteristics through numerical simulations.
Specifically, we introduced an interaction parameter $Q$ into a 3D-KI model to set the relative contributions of metric and topological interactions, enabling us to simultaneously observe the collective behaviour of particles influenced by both interactions.

We quantitatively analyzed a hallmark of topological interactions---the formation of multiple spatial clusters---by applying the HDBSCAN algorithm.
Even in regimes where the global order parameter calculated for all particles was low, the order parameter recomputed for each HDBSCAN-identified cluster remained close to 0.9, indicating a high degree of internal alignment within every cluster.
Thus, this approach enables us to determine in a fully quantitative manner whether the particle ensemble self-organizes into coherent local clusters.
By combining the conventional global order parameter with the HDBSCAN-based clustering framework introduced here, we achieved a more accurate, three-dimensional characterization of collective particle dynamics.

Moreover, our model enables metric and topological interactions to operate simultaneously, a behavior that cannot be reproduced by conventional interaction-switching frameworks. 
We introduced this capability to reflect the possibility that real animals may employ deliberately ambiguous decision rules when forming and maintaining group structures.
At low densities, topological interactions allow particles to establish networks with distant neighbors; however, because interparticle distances are large, a cluster that forms and stabilizes tends to involve the same set of partners, leading to the decline in the global order parameter observed in Fig. \ref{fig3}.
At high densities, the number of nearby particles increases, and an ultra-local network emerges. 
The slightly smaller variance of the order parameter compared with the metric case (Fig. \ref{fig4}) indicates that the system consolidates into a more stable, single flock.
However, because metric interactions act on all neighbors within a spherical region of radius $r$, regardless of neighbor count, they are more effective in forming a flock in low- and intermediate-density regimes.
Consistently, Fig. \ref{fig3} shows that the metric model is more robust to density fluctuations than the topological model.
However, at very low densities, particles may have no neighbors inside the interaction radius, and the additional distance-decay factor prevents collective formation.
When the topological model is blended with a metric component ($Q = 0.75$), each particle operates in an environment where many neighbors lie within both the rank and metric ranges, allowing it to track a larger set of particles continuously. 
Conversely, adding a topological component to the metric model ($Q = 0.25$) helps maintain the particle-to-particle network as the density decreases and the number of metric neighbors declines.
In this framework, metric interactions mainly supply proximity information, whereas topological interactions maintain the network structure.
As a result, Figs. \ref{fig3} and \ref{fig4} reveal a parabolic-shaped boundary in the dependence of the order parameter $v_a$ on the interaction parameter $Q$ and density $\rho$, with the apex located near $Q\approx0.5$.
Thus, at approximately $Q=0.5$, the two interaction types complement each other most effectively, enabling the system to sustain a single coherent flock even when the density is substantially reduced.
These findings demonstrate that allowing metric and topological interactions to act concurrently enhances model robustness (for other noise parameters, see Supplementary Fig S2-S5). 

Nevertheless, the current formulation assumes a single-species system of volume-less, constant-speed particles moving under periodic boundary conditions.
Therefore, additional validation is therefore required before the model can be extended to mixed-species assemblages or habitats with finite boundaries.
Behavioral control laws inferred directly from empirical data have already been extracted for a variety of taxa, such as analyses of golden shiners (\textit{Notemigonus crysoleucas}), which showed that speed regulation dominates inter-individual interactions\cite{bib30}, while a model derived from experimental bird data introduced a new empirical framework for characterizing flocking as a collective phenomenon\cite{bib31}.
In future work, we aim to enhance the model’s generality by fitting it to three-dimensional tracking datasets obtained from wild animal groups.

We also envisage that the proposed model will become a versatile control scheme that can be embedded in a wide range of control-engineering systems.
In swarm robotics, there is a constant demand for control laws that remain robust under rapidly changing environmental conditions. 
Numerous studies have adopted related modeling frameworks in control engineering, particularly swarm robotics. Examples include a shepherding (“sheep-dog”) system that steers a flock toward a target\cite{bib14}, an approach inspired by smoothed-particle hydrodynamics that evaluates density-field interactions through both simulation and physical multi-robot experiments\cite{bib33}, and an experimental study demonstrating that a control algorithm stays stable when deployed on a team of autonomous quadcopters\cite{bib34}.
The next step is to treat the interaction parameter $Q$---introduced in this study as a fixed value within a single simulation---as a state-dependent variable that is switched automatically. 
Currently, $Q$ remains constant throughout each run. 
Based on the trends shown in Figs. \ref{fig3} and \ref{fig4}, however, we can recast $Q$ as a continuously varying function of environmental factors such as density $\rho$ or cluster configuration, thereby transforming it into a robust control rule that adapts to surrounding conditions.
In short, we can broaden the model’s applicability by fitting it to empirical 3D animal-tracking data, while simultaneously refining it into a robust, environmentally adaptive control framework through the dynamic modulation of the interaction parameter $Q$.


\section*{Declaration of competing interest}
The authors declare that they have no known competing financial interests or personal relationships that could have appeared to influence the work reported in this paper.

\section*{Declaration of generative AI and AI-assisted technologies in the writing process}
During the preparation of this manuscript, the authors drafted sections in Japanese and used ChatGPT (OpenAI, San Francisco, CA, USA) to assist with their translation and stylistic refinement in English. 
After using this tool, the authors reviewed and edited the content as needed and take full responsibility for the final content of the manuscript.

\section*{Funding}
This work was supported by JST SPRING, Grant Number JPMJSP2132.

\section*{Data availability}
3D-KI model is developed under processing 4.3.4, and HDBSCAN is performed under Python 3.9.6.
Each codes will be distributed after publication.

\section*{CRediT authorship contribution statement}
{\bf Yuto Kikuchi}: Conceptualization, Methodology, Formal analysis, Investigation, Writing - original draft, Visualization.
{\bf Mayuko Iwamoto}: Supervision, Writing - review \& editing, Methodology Conceptualization.

\section*{Acknowledgements}
The authors thank Prof.Yasufumi Yamada of Future University Hakodate for helpful suggestions.
We would like to thank Editage (www.editage.jp) for English language editing.

\appendix
\section{HDBSCAN for clustering}
\label{app1}
HDBSCAN (Hierarchical Density-Based Spatial Clustering of Applications with Noise) is a density-based clustering algorithm that hierarchically extends DBSCAN.
Rather than fixing a single neighborhood radius $\epsilon$, HDBSCAN sweeps $\epsilon$ continuously through the data space, evaluates the stability of the resulting clusters, and finally extracts an “optimal” flat clustering.
This procedure allows clusters with different densities to be identified simultaneously and makes the method markedly less sensitive to parameter tuning than DBSCAN.
The clustering pipeline used in this study is as follows:

1. Compute the pairwise reachability distances and build the minimum-spanning tree (MST).

2. The MST is converted into a condensed dendrogram by progressively contracting the edges as $\epsilon$ increases.

3. Clusters are selected from the condensed dendrogram based on the basis of their stability scores.

Here, the mutual reachability distance between particles $i$ and $j$ is defined as a linear combination of their Euclidean separation (computed under periodic boundary conditions) and directional distance (i.e., the similarity of their orientation vectors).
For a cubic domain of side length $L$, the Euclidean distance $d_{ij,t}^{pos}$ between particles $i$ and $j$ at time $t$ is given by:

\begin{equation}
 d^{pos}_{ij,t}=\|\Delta_{ij,t}\|_2 \ \ (\Delta_{ij,t}\equiv(\Delta_{ij,t}^{x}\ ,\Delta_{ij,t}^y\ ,\Delta_{ij,t}^z))
\end{equation}
\begin{equation}
 \Delta^{(k)}_{ij,t}=\min(|{\bf x}^{(k)}_{i,t}-{\bf x}^{(k)}_{j,t}|,L_k-|{\bf x}^{(k)}_{i,t}-{\bf x}^{(k)}_{j,t}|),\ \ k\in {x,y,z}
\end{equation}
Scaling to the interval $[0,1]$ is performed as follows:
\begin{equation}
  \tilde{d}^{pos}_{ij,t}=\frac{d^{pos}_{ij,t}}{d_{max}}\ ,
\end{equation}
where
\begin{equation}
  d_{max}=\frac{\|L\|_2}{2}
\end{equation}
The directional‐vector similarity is defined as
\begin{equation}
  d^{vec}_{ij,t}=1-\cos{\theta_{ij,t}}
\end{equation}
Scaling to the interval $[0,1]$ is performed as follows:
\begin{equation}
  \tilde{d}^{vec}_{ij,t}=\frac{{d}^{vec}_{ij,t}}{2}
\end{equation}
Using the mixing weight $w$, we first compute the mutual reachability distance between every pair of points as follows:
\begin{equation}
  d_{ij,t}=w\tilde{d}^{pos}_{ij,t}+(1-w)\tilde{d}^{vec}_{ij,t}
\end{equation}
This distance allowed us to define the core distance, which served as a proxy for the local density around each point.
\begin{equation}
  \text{core}_k(i)=d_{i(k)}
\end{equation}
where $d_{i(k)}$ denotes the $k$-th smallest nonzero distance from particle $i$ to any other particle.
To prevent widely separated low-density points from being clustered together, we rescale the core distances. 
The rescaled values yield the mutual reachability metric used by HDBSCAN:
\begin{equation}
  d^{mr}_k(i,j)=\max{\{\text{core}_k(i),\ \text{core}_k(j),\ d_{ij}\}}
\end{equation}
We assigned the mutual-reachability distance $d_{k}^{mr}(i,j)$ as the edge weight $\omega$ on every edge of the complete graph, constructed a single minimum-spanning tree (MST), and successively cut its edges in descending order of length.
After each cut, if the resulting connected component contains at least the minimum number of points, we update these points with a new cluster label; otherwise, the component is marked as noisy.
This procedure was repeated until each point was isolated, yielding a condensed dendrogram.

Next, we set $\lambda=1/\omega$ and recorded, as a hierarchical tree, how the connected components evolved as $\lambda$ increased. 
We track the density contours of the data landscape and decided which peaks should be regarded as individual clusters that should be merged.
To make this decision, we define the stability of a cluster as follows:
\begin{align}
   S(C)&=\int^{\lambda_{\text{death}}}_{\lambda_{\text{birth}}}|C(\lambda)|d\lambda \notag\\
       &=\sum_{p\in C}(\lambda_p-\lambda_{\text{birth}}(C))
\end{align}
For each cluster $C$, we define $\lambda_{\text{birth}}$ as its birth level, $\lambda_{\text{death}}$ as its death level, $|C(\lambda)|$ as the change in cluster size between the two levels, and $\lambda_p$ as the level at which point $p$ leaves the cluster.
As cluster selection candidates, we consider the most stable minimal branches and compare the quantities $\sum_iS(C_i)$ and $S(C)$ between parent cluster $C$ and its child clusters $\{C_i\}$.
If the children’s score exceeded that of the parent, the child clusters were retained, and the parent cluster was discarded; if the parent’s score was higher, the parent was retained, and the children were discarded.
Iterating this comparison upward toward the root yields a set of clusters that maximizes the overall stability. 
Any point whose label is finally set to $-1$ is treated as noise and assigned to no cluster.




  \bibliographystyle{elsarticle-num-names} 
  \bibliography{reference}






\end{document}